\RequirePackage{amsthm}
\documentclass[sn-mathphys,Numbered]{sn-jnl}
\usepackage{graphicx}%
\pdfoutput=1
\usepackage{multirow}%
\usepackage{amsmath,amssymb,amsfonts}%
\usepackage{amsthm}%
\usepackage{mathrsfs}%
\usepackage[title]{appendix}%
\usepackage{xcolor}%
\usepackage{textcomp}%
\usepackage{manyfoot}%
\usepackage{booktabs}%
\usepackage{algorithm}%
\usepackage{algorithmicx}%
\usepackage{algpseudocode}%
\usepackage{listings}%
\usepackage{braket}
\usepackage{hyperref}
\usepackage{bookmark}

\theoremstyle{thmstyleone}%
\theoremstyle{thmstyletwo}%
\theoremstyle{thmstylethree}%
\begin{document}

\title[Article Title]{Geometric inhibition of superflow in single-layer graphene suggests a staggered-flux superconductivity in bilayer and trilayer graphene}

\author[1]{\fnm{Xinyao} \sur{Zhang}}\email{xinyao@sjtu.eu.cn}

\author[1]{\fnm{Ruoshi} \sur{Jiang}}\email{ruoshijiang@sjtu.edu.cn}

\author[1]{\fnm{Xingchen} \sur{Shen}}\email{shenxchen@sjtu.edu.cn}

\author[3]{\fnm{Xiaomo} \sur{Huang}}\email{hxm0003@sjtu.edu.cn}

\author*[1,4]{\fnm{Qing-Dong} \sur{Jiang}}\email{qingdong.jiang@sjtu.edu.cn}

\author*[1,3,4]{\fnm{Wei} \sur{Ku}}\email{weiku@sjtu.edu.cn}

\affil[1]{\orgdiv{Tsung-Dao Lee Institute  \& School of Physics and Astronomy }, \orgname{Shanghai Jiao Tong University}, \orgaddress{\street{Pudong}, \city{Shanghai}, \postcode{201210},
%\state{State},
\country{China}}}

\affil[2]{\orgdiv{Zhiyuan College}, \orgname{Shanghai Jiao Tong University}, \orgaddress{\city{Shanghai}, \postcode{200240},  \country{China}}}

\affil[3]{%\orgdiv{Shanghai Branch},
\orgname{Key Laboratory of Artificial Structures and Quantum Control}, \orgaddress{\city{Shanghai}, \postcode{200240}, \country{China}}}

\affil[4]{\orgdiv{Shanghai Branch}, \orgname{Hefei National Laboratory}, \orgaddress{\city{Shanghai}, \postcode{201315}, \country{China}}}

\abstract{
In great contrast to the numerous discoveries of superconductivity in layer-stacked graphene systems, the absence of superconductivity in the simplest and cleanest monolayer graphene remains a big puzzle.
Here, through realistic computation of electronic structure, we identify a systematic trend that superconductivity appears to emerge only upon alteration of the low-energy electronic lattice from the underlying honeycomb atomic structure.
We then demonstrate that this inhibition can result from from geometric frustration of the bond lattice that disables quantum phase coherence of the order parameter residing on it.
In comparison, upon deviating from the honeycomb lattice, relief of geometric frustration allows robust superfluidity with non-trivial spatial structure.
For the specific examples of bilayer and trilayer graphene under an external electric field, such bond centered order parameter would develop superfluidity with staggered flux that breaks the time-reversal symmetry.
Our study also suggests the possible realization of the long-sought superconductivity in single-layer graphene via the application of uni-directional strain.
}


\maketitle

\section{Introduction}\label{sec1}

Recently, the discovery of superconductivity in Bernal bilayer graphene (BLG) \cite{Zhou_AB_2022} and rhombohedral trilayer graphene (TLG) \cite{Zhou_ABC_2021} attracted great research interest.
The discovered superconductivity exhibits an extremely large upper critical field that exceeds the Pauli paramagnetic limit.
Additionally, the normal-state magnetoresistance shows an unusual linear magnetic field dependence~\cite{Kisslinger_2015,Jhang_2011}.
Similar unconventional normal-state and superconducting properties have also been observed in varies moir\'e systems, such as twist bilayer graphene \cite{Cao_TBG_2018,Yankowitz_TBG_2019,Lu_TBG_2019,Saito_TBG_2020,Cao_TBG_strange_2020}, twist trilayer graphene \cite{Chen_ABC1_2019,Chen_ABC2_2019,Cao_TTG_2021,Tsai_TTG_2021}, twist multi-layer (N$>$3) graphene \cite{Cao_mutiLG_2022}, and twist bilayer $\rm WSe_2$ \cite{Wang_2020,Ghiotto_2021}.
In fact, similar unusual properties are also observed in various monolayer transition metal dichalcogenides  superconductors, including $\rm WTe_2$ \cite{Sajadi_2018}, $\rm NbSe_2$ \cite{Ichinokura_2019}, and $\rm MoS_2$ \cite{Fu_2017}.
These rather exotic properties pose a significant challenge to the traditional picture of transport and superconducting behaviors, and imply significant research potential in exploring new states of matter in these two-dimensional materials beyond the Fermi liquid paradigm.

Strangely, despite the enormous amount of experimental efforts, the cleanest single-layer graphene (SLG) has not shown any sign of superconductivity.
In contrast to the alkali metal intercalated graphene families \cite{Chapman_SLG(Li)_2015,Ichinokura_SLG(Ca)_2016,Ludbrook_SLG(Li)_2015} that display superconductivity, inserting a large amount of carrier via gating does not seem to induce superconductivity \cite{Chen_2008,Saha_unconventional_2014}.
This suggests that some fundamental physical reason beyond carrier density is at play that prohibits the emergence of superconductivity in SLG.

Here, we address this important issue by seeking answers to three specific questions:
1) Is there a qualitative trait in the electronic structure that correlates with the emergence or absence of superconductivity?
2) Is there a simple physical mechanism that can explain such a trait, particularly the inhibition of superconductivity in SLG?
and
3) According to such a theory, are there any unusual superconducting properties in, for example, the BLG and TLG that can be experimentally verified?

\section{Lattice geometry of low-energy electrons}\label{sec2}
\begin{figure}[htp]%
\centering
\includegraphics[width=0.9\textwidth]{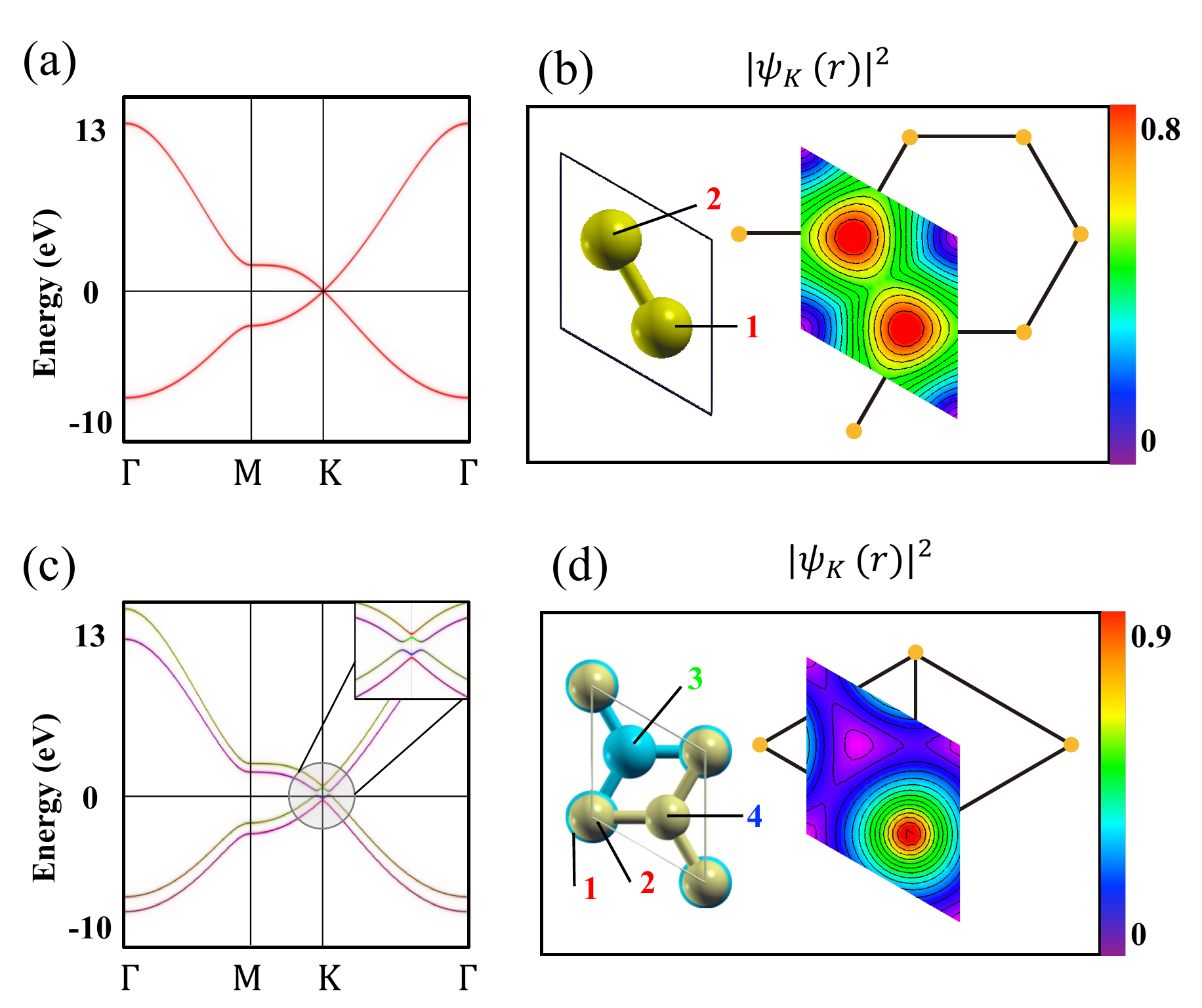}
\caption{Low-energy electronic lattice of SLG and BLG. The  $p_z$ band dispersion of SLG in (a) corresponds to low-energy Bloch orbitals in the proximity of Fermi level in (b).
    (c) and (d) give the same for BLG, colored by red, green, and blue to represent components from orbitals in atoms 1, 2, 3, and 4 in the unit cell shown in (d).
    Notice that the low-energy electrons in SLG reside in the same honeycomb lattice as the underlying atomic lattice, while in BLG, they reside in a triangular lattice instead.}
\label{fig1}
\end{figure}

To seek a qualitative trait in the electronic structure that correlates with the emergence or absence of superconductivity, we perform density functional calculation \cite{Hohenberg_1964,Kohn_1965} for BLG under a large perpendicular field, and compare the low-energy orbitals near the chemical potential with those in SLG.
Figure \ref{fig1} shows the low-energy band structure for SLG (a) and BLG (c) under large a perpendicular electric field, and the corresponding spatial distribution of the Bloch orbitals (b) and (d) slightly below the chemical potential (chosen as zero energy), in which the doped hole-carriers in these systems reside.

Interestingly, Fig. \ref{fig1} (b) and (d) show a qualitative difference in the spatial distribution of the low-energy orbitals in these two materials, despite their similar band dispersion (a) and (c).
The electron carriers in SLG propagate at the original honeycomb lattice of the atoms, but the hole carriers in BLG reside only in a \textit{triangular} lattice corresponding to one of the ''AB'' sites.
(This is easily understood from the external perpendicular field that breaks the symmetry connecting the two AB sites within the graphene unit cell.)

In fact, such an alteration of the electronic lattice from the underlying honeycomb atomic lattice, for example to a triangular lattice in BLG, takes place in \textit{all} graphene-derived materials that demonstrate superconductivity, such as BLG, TLG, and twisted bilayer graphene.
Consistently, the triangular electronic lattice in the superconducting transition metal dichalcogenides \cite{Shi_TMD_2015,Saito_ZrNCl_2015,Zeng_SnSe2_2018,Trainer_proximity_2020} and the Kagome electronic lattice in the superconducting Kagome metals \cite{Ortiz_2020,Yin_2021,Jiang_2021} also do not have a honeycomb electronic lattice.
On the other hand, to the best of our knowledge, so far none of the graphene-derived materials that retain the honeycomb electronic lattice, such as SLG and AA stacking bilayer graphene, is able to display superconductivity.
We thus have identified an important qualitative trait, namely the honeycomb geometry of the electronic lattice, that appears to prohibit the emergence of superconductivity.

So, what could be the physical reason for the honeycomb geometry of the electronic lattice to be harmful to superconductivity?
The current BCS framework-based proposals via various fluctuation~\cite{BCS,MacDonald_2018,Liu_2018,Lian_2020,Fischer_2021} are insensitive to the lattice geometry and thus offers no qualitative answer to this question.
In fact, current experiments on these materials already reveal quite a few non-Fermi liquid transport properties in the `normal state' above the superconducting temperature, including non-saturating high-temperature resistivity (so-called “bad metal'') \cite{Eszter_2003}, linear temperature-dependent resistivity (“strange metal'') \cite{Chen_2008,Dean_2010,Saha_unconventional_2014,Cao_TBG_strange_2020,Zhou_ABC_2021,Zhou_AB_2022,Cao_mutiLG_2022}, and insulator-like enhancement of low-temperature resistivity (“weak insulator'')\cite{Chen_2008,Heo_nonmonotonic_2011,Cao_TBG_strange_2020,Zhou_ABC_2021,Zhou_AB_2022,Cao_mutiLG_2022}.
Therefore, one really should seek alternative theories beyond Fermi liquid that offers a physically consistent description of the higher-temperature normal state, before investigating the emergence (or absence) of the lower-energy superconductivity.

\section{Simple picture via emergent Bose liquid}\label{sec3}

%\section{Low-energy effective structure}\label{sec3}
To this end, below we will explore the simplest non-Fermi liquid theory, the emergent Bose liquid (EBL) theory, whose \textit{intrinsic} normal-state properties coincide with all these non-Fermi liquid behaviors, including exotic transport properties \cite{Jiang_2017,Zeng_2021,Yue_2023} and one-body spectral functions \cite{Jiang_2017,Zeng_2021,Yue_2023}.
Given that nearly all phenomenological behaviors of these materials are automatically captured by the intrinsic characteristics of EBL, this theory offers a simple and natural starting point to address the questions at hand.
Particularly, EBL, like all interacting bosonic systems, has a nearly inescapable tendency toward superfluidity \cite{Wei_2011,Wei_2015,Lang_2022} at low temperature.
Therefore, if even in such a superfluid-favoring system a geometry-sensitive mechanism can be identified that prohibits its default low-temperature state of superfluidity, such mechanism must have a rather general applicability beyond a specific model.

The conditions resulting in an EBL \cite{Wei_2011,Wei_2015,Lang_2022,Hegg_2021} are quite general and begin with strong local physics that prohibits double occupation of carriers in each atom, such as strong intra-atomic repulsion or, in the case of graphene, Pauli principles in weakly doped half-filled systems.
In such highly constrained systems, strong short-range correlations, for example, anti-ferromagnetic correlation \cite{Anderson_1952} and bi-polaronic correlation \cite{Alexandrov,Jun_1992}, are expected to emerge at rather a high-energy scale (using the kinetic energy as a reference).
If we then assume that from these correlations an effective binding mechanism emerges, the physics will be dominated by bosons formed from the \textit{nearest neighboring} carriers.
These bosons have centers at the \textit{bonds} of the underlying fermion lattice and are spatially separated due to their extended hardcore nature (forbidding occupation of their surrounding bounds) inherited from the above intra-atomic high-energy constraint.

Note that the specific details of the binding mechanism are actually not very important, since lower-energy physics are unable to overcome the binding scale, and are therefore \textit{completely insensitive} to it.
This is analogous to the well-known example of an ideal gas consisting of H$_2$ molecules, whose properties at room temperature and ambient pressure are completely insensitive to the higher-energy microscopic mechanism that binds the two H atoms into a molecule.

As illustrated below, an essential feature of an EBL is the multi-orbital nature of the bond lattice it resides, in association with multiple bond orientations in 2D and 3D systems.
Such multi-orbital nature makes EBL generically \textit{sensitive to the local geometry}.
This has recently led to the realization of a novel homogeneous quantum Bose metal state \cite{Hegg_2021} resembling the pseudogap phase \cite{Yue_2023} found in several strongly correlated materials.

\begin{figure}[htp]%
\centering
\includegraphics[width=0.8\textwidth]{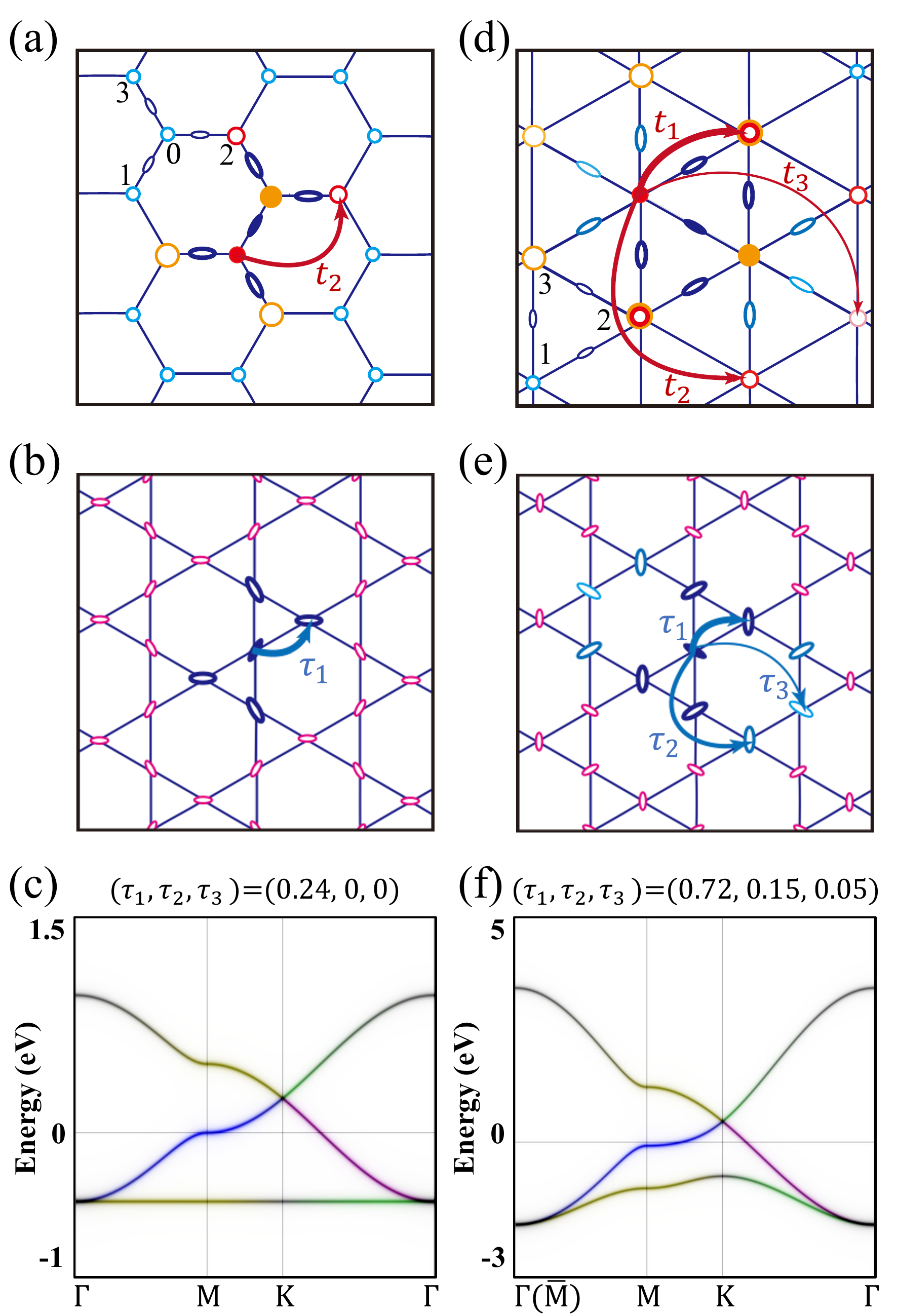}
\caption{Construction of emergent Bose liquid and resulting geometric frustration.
    For fermionic carriers on a honeycomb lattice shown in (a), the emergent bosons reside on the Kagome bonds lattice in (b).
    Upon formation of bosonic bound states, the fermionic near-neighboring hoppings $t_1$ and $t_3$ are suppressed, leaving only the nearest neighboring hopping $\tau_1=t_2=0.24$ eV (which preserves the bound state) in the kinetic process of EBL.
    Due to the perfect geometric frustration of the Kagome lattice, The resulting kinetic dispersion in (c) has a completely flat band.
    (d)(e)\&(f) give the same for a triangular lattice.
    Even though the bosonic bond lattice is also a Kagome one, the different orientation of the bonds allows some $t_1$, $t_2$, and $t_3$ to preserve the bound states in (e).    Consequently, the kinetic dispersion in (c) no long suffer from geometric frustration.}
    \label{fig2}
\end{figure}

Figure \ref{fig2}(a) illustrates the construction of the EBL model on a honeycomb fermionic lattice.
Consider a generic effective fermionic Hamiltonian, for example,
\begin{equation}
H^\text{F}=\sum_{ii'} t_{ii'}c^{\dagger}_ic_{i'}+\sum_{ii'}U_{ii'}c^{\dagger}_ic^{\dagger}_{i'}c_{i'}c_i,
\end{equation}
describing the kinetic strength $t_{ii'}$ and two-body interaction strength $U_{ii'}$ of the carriers, where $c^\dagger_i$ denotes creation of an itinerant fermionic carrier on lattice sites $i$.
Following the above assumptions of EBL formation, one starts by first forming a bond lattice with boson $a^\dagger_j = c^\dagger_i c^\dagger_{i^\prime}$ located at the \textit{bond} $j$ connecting the fermionic lattice sites $i$ and $i^\prime$.
As shown in Fig. \ref{fig2}(b), the bond lattice of the honeycomb lattice is instead a \textit{Kagome} lattice, on which the emergent bosons reside.

Next, the leading \textit{kinetic} process of the bosonic carriers,
\begin{equation}
H^\text{B}=\sum_{jj'} \tau_{jj'}a^{\dagger}_ja_{j'},
\end{equation}
can be obtained from
\begin{equation}
 \tau_{jj'}=\bra{0}a_jH^\text{F} a^\dagger_{j'}\ket{0},
 \label{tau}
\end{equation}
since higher-order kinetic processes are suppressed by the $t_{ii'}/E_\text{binding}$ factor (multiple times).
For this illustration, we use fermionic hopping strengths $t$'s obtained from a Wannier function analysis \cite{Wei_2002} of the first-principles electronic band structure shown in Fig. \ref{fig1}.
For simplicity, we choose the symmetric convention $a^\dagger_j\equiv c^\dagger_0c^\dagger_1$, $a^\dagger_j\equiv c^\dagger_1c^\dagger_2$, and $a^\dagger_j\equiv c^\dagger_2c^\dagger_0=-c^\dagger_0c^\dagger_2$, for sites $i=0,1,\text{ and}\ 2$ shown in Fig. \ref{fig2}(a).

Most significantly, from Fig.~\ref{fig2}(a) and (b) one can observe that Eq. \ref{tau} generates \textit{only} the nearest neighboring bosonic hopping $\tau_1=t_2$ from the underlying next-nearest neighboring fermionic hopping $t_2=0.24$ eV.
Other fermionic kinetic terms, such as the nearest neighboring fermionic hopping $t_1$ and third-neighboring hopping $t_3$ both break apart the boson and thus should be “integrated out'' from the low-energy sector, unable to contribute directly through Eq. \ref{tau}.

\section{Inhibition of superfluidity due to geometric frustration}\label{sec4}

Remarkably, the straightforward application of the EBL model offers a natural explanation for the inhibition of superconductivity in SLG.
Figure~\ref{fig2} (c) shows the resulting kinetic energy dispersion from $H^\text{B}$ derived above, with the \textit{only} allowed $\tau=0.24$ eV in the Kagome bosonic lattice.
In contrast to the conventional bosonic systems characterized by a single minimum in their kinetic dispersion, the EBL model described here gives rise to an entirely flat dispersion at the bottom of the band structure.
Such a completely flat dispersion reflects the existence of a large number (the number of system lattice sites) of compact localized one-body orbitals \cite{Read_2017} in such a geometrically frustrated system.
This result resembles exactly the recent discovery~\cite{Hegg_2021} of quantum Bose metal state in geometrically frustrated system that \textit{defeats} the superfluid state by disabling the quantum coherence associated with the kinetic processes.
In fact, this lack of coherence can be rigorously proven~\cite{Zhang23} to persist even in the presence of typical Coulomb interaction, and thus completely rules out the possibility of inducing superfluidity via interaction~\cite{You_2012,Julku_2021}.
 
In other words, even though the non-Fermi liquid behavior inspires our exploration of the EBL model, against bosons' nearly inescapable nature toward superfluidity, the geometric frustration still manages to inhibit its superconductivity!
In essence, the perfect \textit{geometric frustration} of the bosonic bond lattice (that emerges from the fermionic honeycomb lattice) destroys the phase coherence of the ground state and disables any potential superfluidity.
Notice that, even though the above analysis is illustrated with the simple EBL model, the consideration of geometric frustration and its inhibition of superfluidity is obviously much general, beyond any specific model.
It should apply to all systems in which order parameters reside on the bond lattice and suffer from geometric frustration.

Naturally, if the above perfect geometric frustration can be removed, a single minimum would develop in the kinetic dispersion and bosonic systems' generic tendency toward superfluidity would prevail again.
Interesting, this expectation is in excellent agreement with the observation of superconductivity in all the graphene-derived superconducting materials, such as BLG, TLG, Ca/Li intercalated SLG, or in twisted bilayer graphene, in all of which the low-energy electronic structure \textit{deviates} from the original honeycomb lattice.
One can therefore make an immediate \textit{prediction} that even single-layer graphene may develop superconductivity (at rather low temperature though) if the symmetry of the honeycomb lattice can be lowered, for example via a uni-axial strain, that breaks the perfect geometric frustration in the EBL.

\section{Multi-component superfluidity}\label{sec5}

As a contrast, let us consider the representative triangular lattice that emerges in the low-energy electronic structure in many of these superconducting materials mentioned above.
Figure \ref{fig2} (d) and (e) show that even though the corresponding bond lattice of EBL is also a Kagome lattice, \textit{three} types of near neighboring hopping processes of the fermionic carriers, through $t_1$ $t_2$, and $t_3$, preserve the binding of the emergent boson.
The resulting EBL thus exercises pivoting motion through bosonic hopping $\tau_1=-t_1$, $\tau_2=t_2$, and $\tau_3=-t_3$.
(Here a similar symmetric convention, $a^{\dagger}_{12}=c^\dagger_2c^\dagger_1$, $a^{\dagger}_{13}=c^\dagger_1c^\dagger_3$, and $a^{\dagger}_{32}=c^\dagger_3c^\dagger_2$, is used.)

Importantly, Fig. \ref{fig2} (f) exemplifies that, as long as $t_2$ and $t_3$ are non-negligible, the bosonic bond lattice generally hosts no perfect geometric frustration, as evidenced by the absence of flat dispersion.
Therefore, an EBL can easily host superfluid phases at low-energy temperatures, as observed experimentally in the above-mentioned materials.

\begin{figure}[htp]%
\centering
\includegraphics[width=0.9\textwidth]{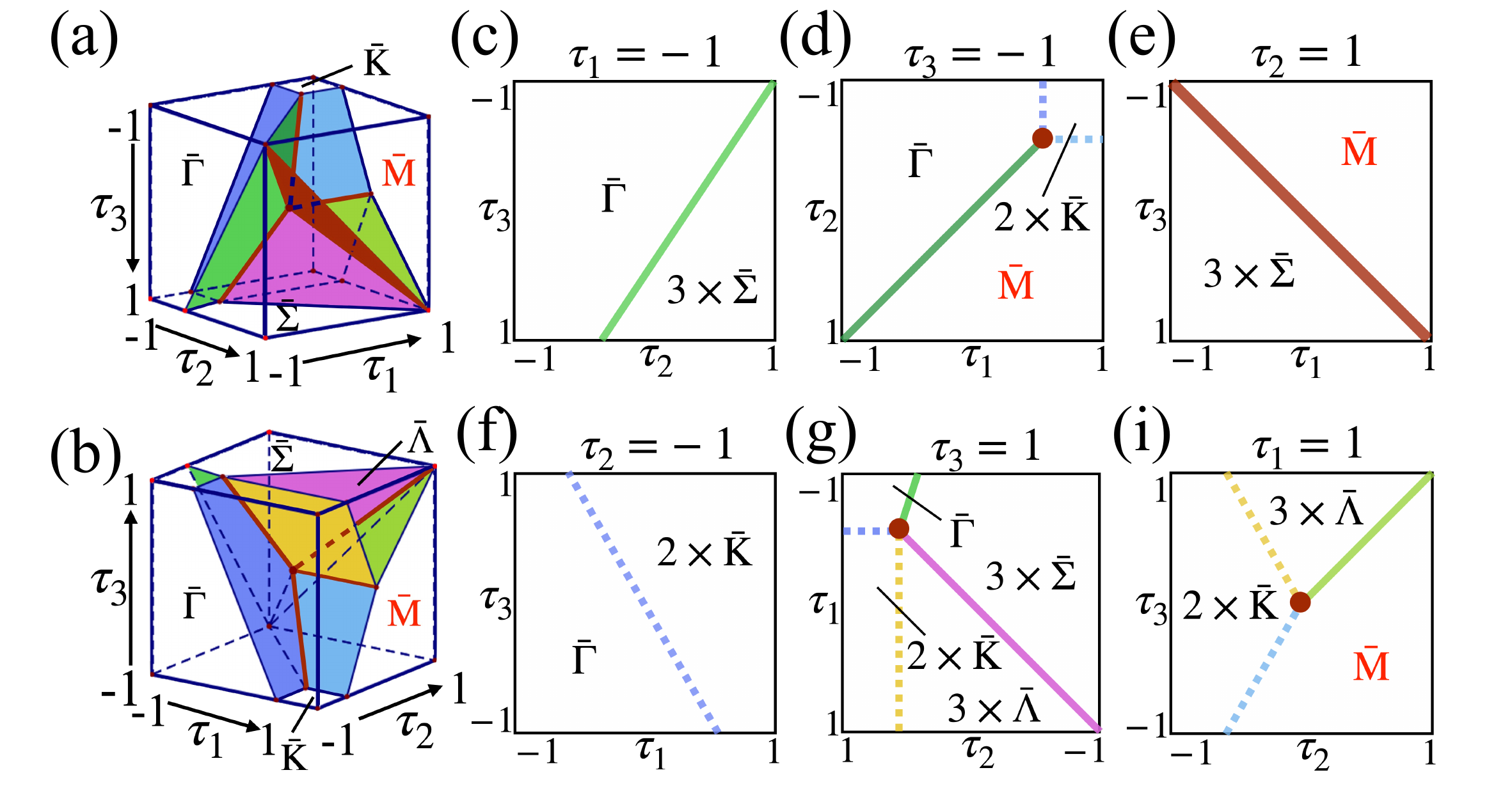}
\caption{Phase diagram of emergent Bose liquid from a triangular fermionic lattice as a function of bosonic kinetic strength $\tau_1$, and $\tau_3$, viewed in 3D (a)\&(b) from different angles and in 2D (c)-(i) on various surfaces of (a).
    The diagrams show five different phases, each labeled with their representative momentum in “unfolded'' indices: a single-component superfluid $\bar{\Gamma}=(0,0)$, a two-component superfluid $\bar{\textrm{K}}=(\frac{2\pi}{3},\frac{2\pi}{3})$, two three-component superfluid $\bar{\Lambda}=(\pi,\pi)$ and $\bar{\Sigma}=(\pi,0)$, and finally a doubly degenerate staggered-flux superfluidity $\bar{\textrm{M}}=(2\pi,0)$ that breaks time-reversal symmetry.
    Second-order and first-order phase boundaries between these phases are represented by solid and dotted lines, respectively.
    The thick dark red lines and surfaces represent boundaries with perfect geometric frustration in the ground state (with flat dispersion).
}\label{fig3}
\end{figure}

Even more interestingly, it turns out several of these superfluid phases have \textit{multi-components} superfluid.
Figure \ref{fig3} gives the full low-temperature phase diagram of EBL for systems hosting triangular fermionic lattice by adjusting the bosonic hoppings.
It shows five possible superfluid phases: a standard zero momentum superfluid at $\bar{\Gamma}$, a two-component superfluid at $\bar{\rm K}$, two three-component ones at $\bar{\Lambda}$ and $\bar{\Sigma}$, and a two-fold degenerate fluid at $\bar{\rm M}$.
These phases are separated by phase boundaries that are either second-order (solid lines) or first-order (dotted lines) in nature, depending on whether the kinetic energy barriers between the minima vanish at the phase boundaries (or not).
For the multi-components superfluid phases, the population of each component can in principle be different, giving rise to additional spatial symmetry breakings, such as a nematic superfluid phase \cite{Shen23} or a chiral superfluid phase \cite{Joseph_1991,Volovik_2009,Jiang_2020}.
 
\begin{figure}[htp]%
\centering
\includegraphics[width=0.9\textwidth]{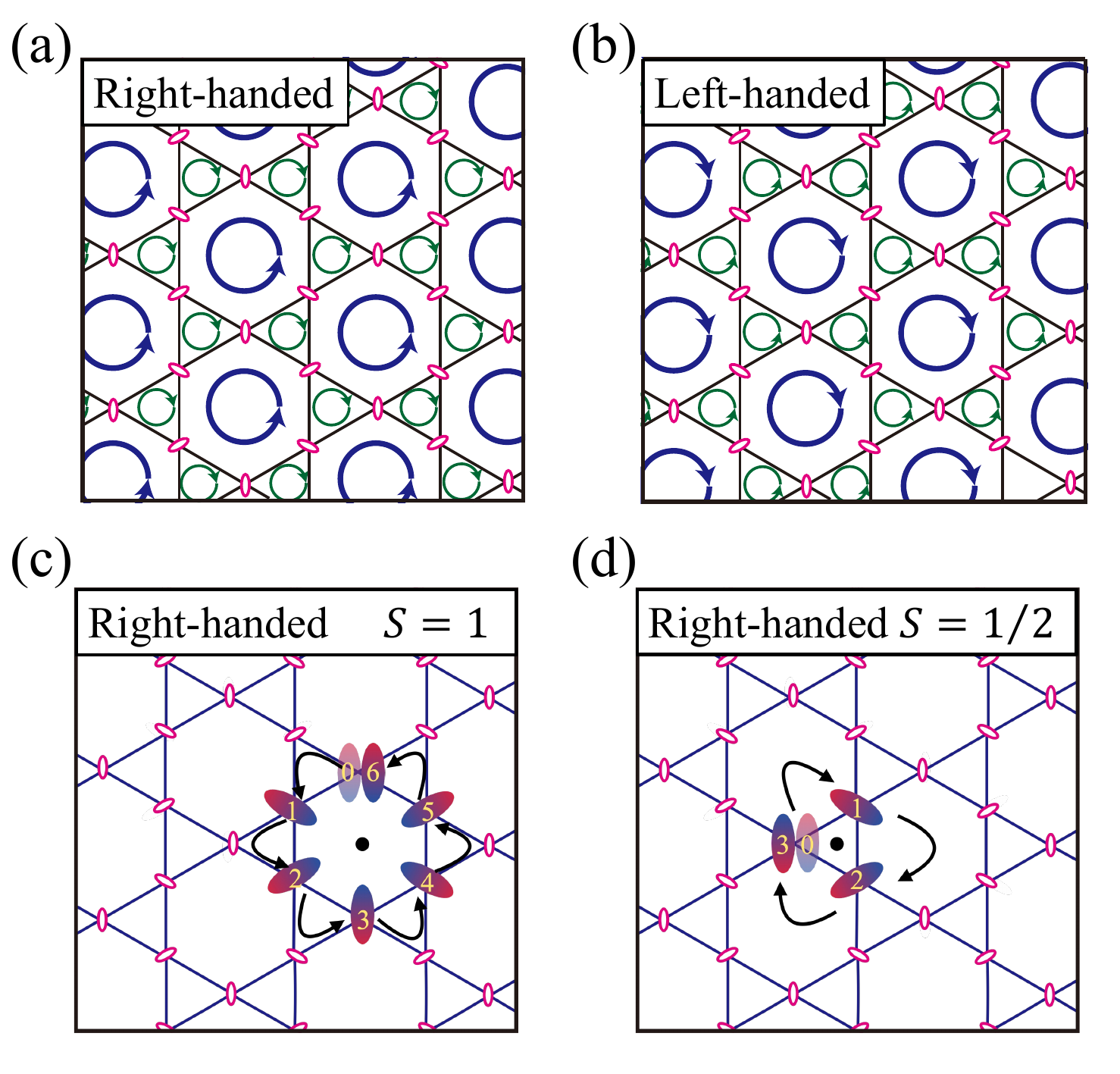}
\caption{(a) and (b) show the degenerate staggered flux (SF) superfluid states.
    Each green orbital current around the triangles has a $2\pi$ flux, countering the $4\pi$ flux of the blue current around the hexagons in the opposite direction.
    However, (c) and (d) shows that all the orbital currents are associated with spinning, $S$, of the emergent boson along the \textit{same} direction, resulting in breaking of time-reversal symmetry.
}\label{fig4}
\end{figure}

Of particular interest is the doubly-degenerate $\rm{\bar{M}}$ superfluid phase, corresponding to a large parameter region including $(\tau_1,\tau_2,\tau_3)\propto(1,1,-1)$.
Our calculated parameters $(\tau_1,\tau_2,\tau_3)$ for BLG and TLG are $(0.72, 0.16,0.05)$ eV and $(0.7, 0.15, 0.04)$ eV, respectively, both deep inside the $\rm {\bar{M}}$ phase.
This phase has a two-fold degeneracy at the same $\rm{\bar{M}}$ momentum, as shown in Fig. \ref{fig2} (f).
It therefore has a strong tendency toward further lowering the symmetry and splitting the degeneracy.
This would give rise to a superfluid phase with staggered magnetic flux \cite{Affleck_flux,Zhang_staggered_1990,Ivanov_2003} with right-handed or left-handed chirality, as illustrated in Fig.~\ref{fig4} (a) and (b).
While the flux associated with the orbital current around the hexagon appears to cancel with that around the two triangles, Fig.~\ref{fig4} (c) and (d) illustrate that both currents also induce effective “spin'' of the emergent boson along \textit{the same} direction.
These states therefore break global time-reversal symmetry, allowing experimental verification for this exotic superfluid phase.
It would be extremely interesting to carefully examine the superconductivity in BLG and TLG for signs of staggered flux superfluidity with weakly broken time-reversal symmetry.

\section{Conclusion}\label{sec6}

In conclusion, we identify the geometry of the honeycomb lattice of the realistic low-energy electronic structure as a key factor in inhibiting superconductivity in graphene-derived materials.
Using emergent Bose liquid as an extreme example, we demonstrate that the lack of superconductivity can be attributed to complete loss of phase coherence due to a perfect geometric frustration in the corresponding bond-centered bosonic lattice.
This suggests a potential avenue for inducing superconductivity by alleviating the perfect geometric frustration in single-layer graphene, for example through application of uni-axial tension.
The geometric consideration also suggests various possibilities of multi-component superconductivity.
Particularly, the observed superconducting phase in the BLG and TLG might correspond to a stagger-flux superconducting state with broken time-reversal symmetry.

\section{Methods}\label{sec7}

\subsection{Visualizing multi-component superfluidity via (unfolded) band structures}\label{subsec1}
\begin{figure}[htp]%
\centering
\includegraphics[width=0.8\textwidth]{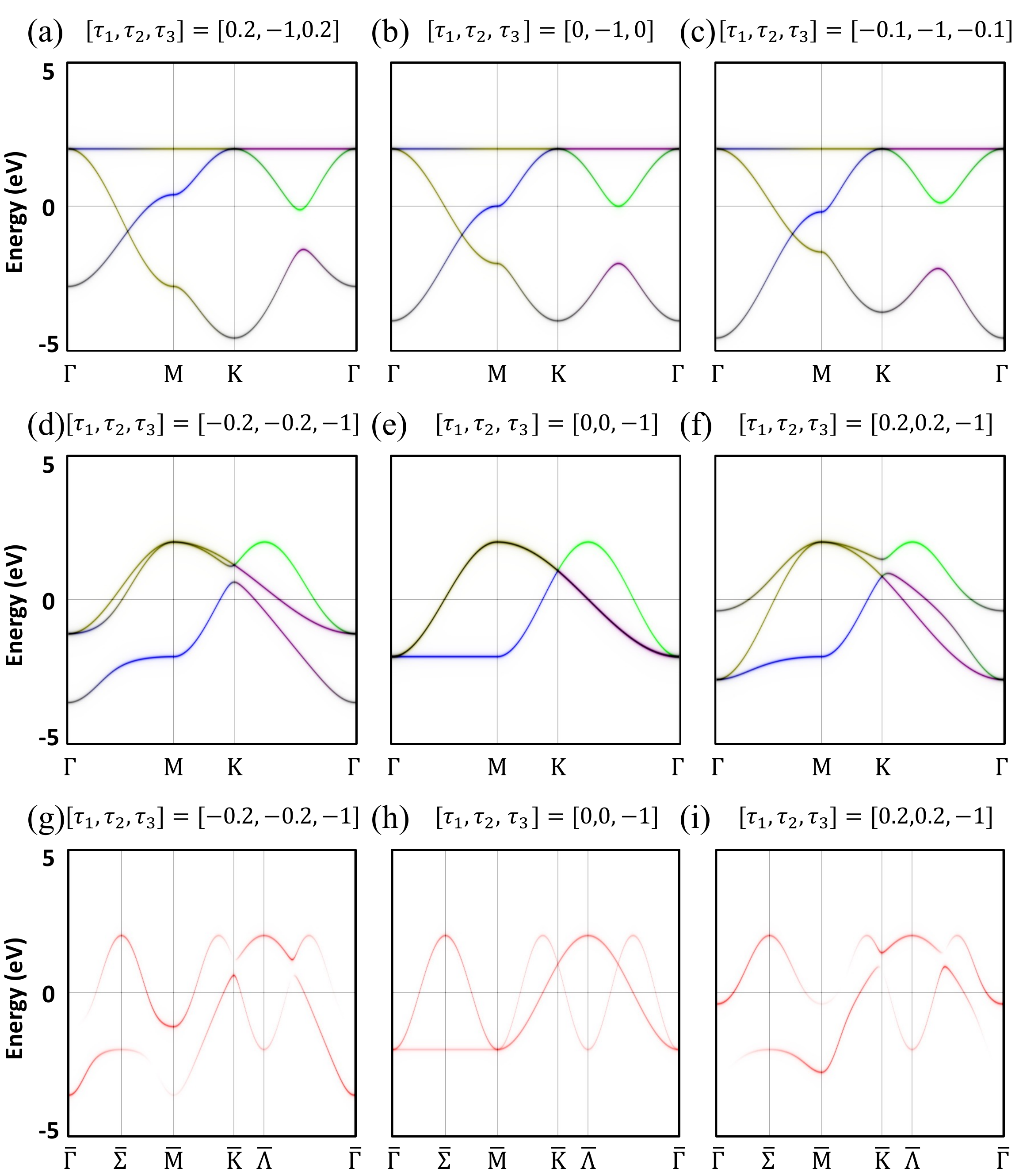}
\caption{Visualizing multi-component superfluidity in Kagome lattice, via EBL band structures under various parameters.
The minima of the band dispersion give possible condensation at the corresponding momenta.
Red, green, and blue colors reflects the strength of the orbitals corresponding to the three bonds in the original triangular fermionic lattice.
(a)-(f) show a first-order switch of superfluidity from a two-component one with momenta $K$ to a single component one with momentum 0.
(d)-(f) show a continuous transition of from a single-component superfluidity with momentum 0 to a two-component one with seemingly similar momentum 0.
The corresponding unfolded band structures in (g)-(i) clarify the latter make of different spatial structures corresponding to momenta $\bar{K}$. 
The high symmetry point positions in the Brillouin zone have coordinates: $\Gamma-(0,0)$, $\Sigma-(\frac{\pi}{2},0)$, $\rm M-(\pi,0)$, $\rm K -(\frac{2\pi}{3},\frac{2\pi}{3})$, and $\rm \Lambda-(\frac{\pi}{2},\frac{\pi}{2})$, in the corresponding Brillouin zone.
}\label{Sfig1}
\end{figure}

The bosonic band structure offers a convenient mean to visualize the spatial phase structure of various multi-component superconductivity.
Figure \ref{Sfig1} gives representative unfolded band structures of EBL in Kagome lattice model, corresponding to different parameters.
The minima of the dispersion indicates condensation and superfluidity (upon stiffening the phase with interaction) with its spatial phase structure related to the corresponding momenta.
Panels (a)-(c) illustrate a switch of minima, indicating a first order phase transition between superfluidity of different momenta.
In contrast, panels (d)-(f) show smooth evolution from one set of minima to another, corresponding to continuous phase transition.
To further help visualize the change of the spatial phase structure, panels (g)-(i) unfold~\cite{ku_unfolding_2010} the bosonic band structure in (d)-(f) to a larger Brillouin zone, corresponding to a smaller spatial unit cell.
The change of momenta in the minimum in (g) and (i) clarifies the different spatial phase structures of the phases in (d) and (f).

\subsection{Evaluation of bosonic hopping parameters}\label{subsec2}

\begin{figure}[htp]%
\centering
\includegraphics[width=0.9\textwidth]{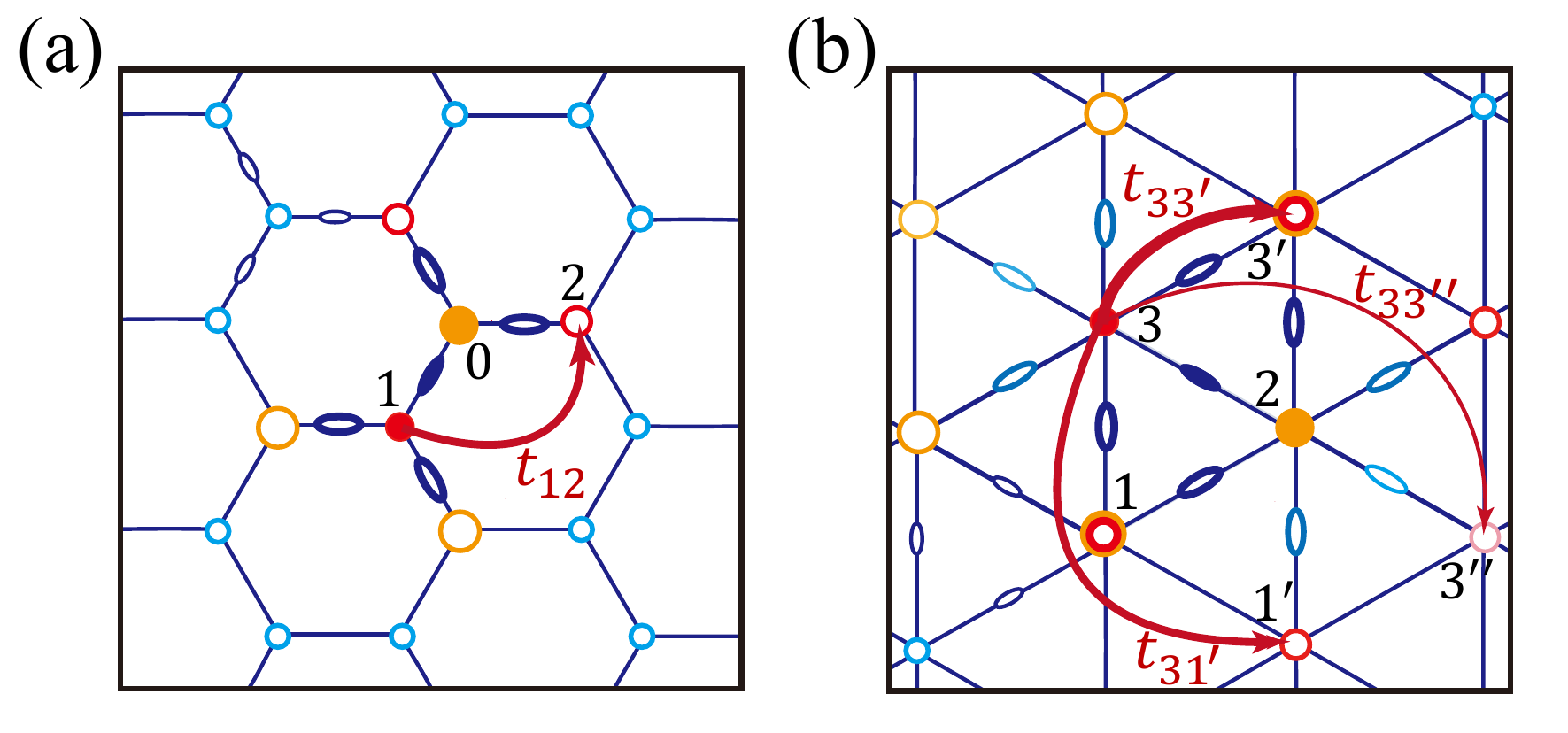}
\caption{ Illustration of the phase convention of bond centered bosons and their hopping parameters inherited from underlying fermionic hoppings in (a) honeycomb lattice and (b) triangular lattice.
}\label{Sfig2}
\end{figure}

Here we give a few examples of evaluation of bosonic hopping parameters according to Eq.~\ref{tau}.
For a Kagome bosonic lattice emerged from honeycomb fermionic lattice with convention defined in Fig.~\ref{Sfig2} (a):
\begin{equation}
\begin{aligned}
   \tau_{1} &=\bra{0}a_jH^\text{F}a^\dagger_{j'_N}\ket{0}\\
   &=\bra{0} c_2 c_0\  t_{12} c^\dagger_2c_1\   c^\dagger_0c^\dagger_1\ket{0}\\
   &=t_{12}=t_{NN}
\end{aligned}
\end{equation}

Similarly, for a Kagome bosonic lattice emerged from triangular fermionic lattice with convention defined in Fig.~\ref{Sfig2} (b):
\begin{equation}
\begin{aligned}
   \tau_1 &=\bra{0}a_jH^\text{F}a^\dagger_{j'_N}\ket{0}\\
   &=\bra{0}c_2c_3\ t_{33'}c^\dagger_{3}c_{3'}\  c^\dagger_{3'}c^\dagger_2\ket{0}\\
   &=-t_{33'}=-t_{N}\\
  \tau_2 &=\bra{0}a_jH^\text{F}a^\dagger_{j'_{NN}}\ket{0}\\
  &=\bra{0}c_2c_3\ t_{31'}c^\dagger_3c_{1'}\ c_2^\dagger c_{1'}^\dagger \ket{0}\\
  &=t_{31'}=t_{NN}\\
   \tau_3 &=\bra{0}a_jH^\text{F}a^\dagger_{j'_{NNN}}\ket{0}\\
   &=\bra{0}c_2c_3\ t_{33''}c^\dagger_3c_{3''}\  c^\dagger_{3''}c_2^\dagger\ket{0}\\
   &=-t_{33''}=-t_{NNN}
\end{aligned}
\end{equation}

\subsection{Estimating the energy scale of strained graphene lattice}\label{subsec3}

\begin{figure}[htp]%
\centering
\includegraphics[width=0.3\textwidth]{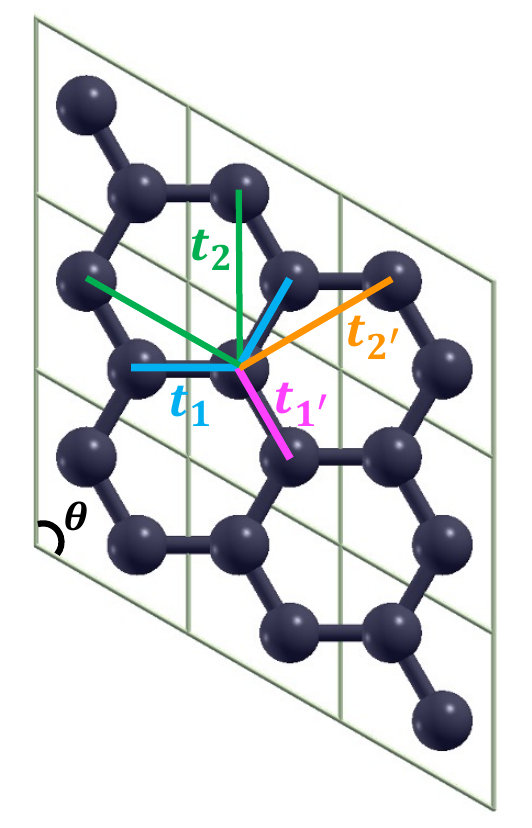}
\caption{ Illustration of lifting geometric frustration via applied uni-axial strain to induce lattice distortion, quantified via deviation of $\theta$ from $120^\circ$. Consequently, the symmetry of the kinetic strength is lowered, $t_1\neq t_{1'}$ and $t_2\neq t_{2'}$.}
\label{Sfig3}
\end{figure}

Here we estimate the energy scale of the induced bandwidth of the geometrically frustrated bosonic flat band in Fig.~\ref{fig2}(c), corresponding to a single-layer graphene under a uni-axial strain quantified by deviation $\delta\theta$ of the angles between unit cell vectors away from $120^\circ$, as shown in Fig.~\ref{Sfig3}.
Table~\ref{tabs1} gives the resulting nearest neighboring fermionic hopping parameters from density functional calculation and Wannier function analysis.
Naturally, $t_1\neq t_{1'}$ and $t_2\neq t_{2'}$ in strained systems since the symmetry is lowered.
Table~\ref{tabs1} shows that the resulting bandwidth of the bosonic flat band of grows $\sim100$ meV per $2^\circ$ of deformation, large enough to host experimentally detectable superconductivity.

\begin{table}[htp]%
\caption{Deformation $\Delta\theta$-dependence of the fermionic nearest and next nearest neighboring hopping parameters, $t_1\neq t_{1'}$ and $t_2\neq t_{2'}$, and the resulting bandwidth, $\Delta W$ of the geometrically frustrated flat band of the corresponding EBL.}
\label{tabs1}%
\begin{tabular}{@{}llll@{}}
\toprule
Distorted angle ($\Delta \theta^\circ$)       & $0^\circ$ & $2^\circ$ & $4^\circ$ \\
\midrule
$t_1$ (eV)    & -2.835  & -2.94  &-2.95  \\
$t_{1'}$ (eV) &  $t_1$    & -2.91 &-2.89  \\
$t_2$ (eV)    & 0.238   & 0.255 & 0.28 \\
$t_{2'}$ (eV) &   $t_2$     & 0.168 & 0.098 \\
\midrule
$\Delta W$ (eV)   & 0      &0.113  &0.233\\
\botrule
\end{tabular}
\end{table}

\backmatter

%\bmhead{Supplementary information}
%If your article has accompanying supplementary file/s please state so here. 

\bmhead{Acknowledgments}
We thank Zi-Jian Lang and Anthony Hegg for useful discussions.
This work is supported by National Natural Science Foundation of China (NSFC) under Grants No.12274287 and 12042507, and Innovation Program for Quantum Science and Technology No. 2021ZD0301900.
Qing-Dong Jiang acknowledge supported from Jiaoda 2030 program, Pujiang Talent Program 21PJ1405400.

\begin{appendices}
\section{Bipolarons as an example of EBL}

Our main conclusions employ only on geometry-sensitive bond-centered order parameters, not necessarily the EBL model used in the study as a simple and extreme example.
Nonetheless, given that EBL is the simplest non-Fermi liquid and that its generic properties appear to coincide with many observed non-Fermi liquid behaviors of graphene-derived materials, it is still interesting to consider the possible formation of EBL in real materials.

\begin{figure}[htp]%
\centering
\includegraphics[width=0.9\textwidth]{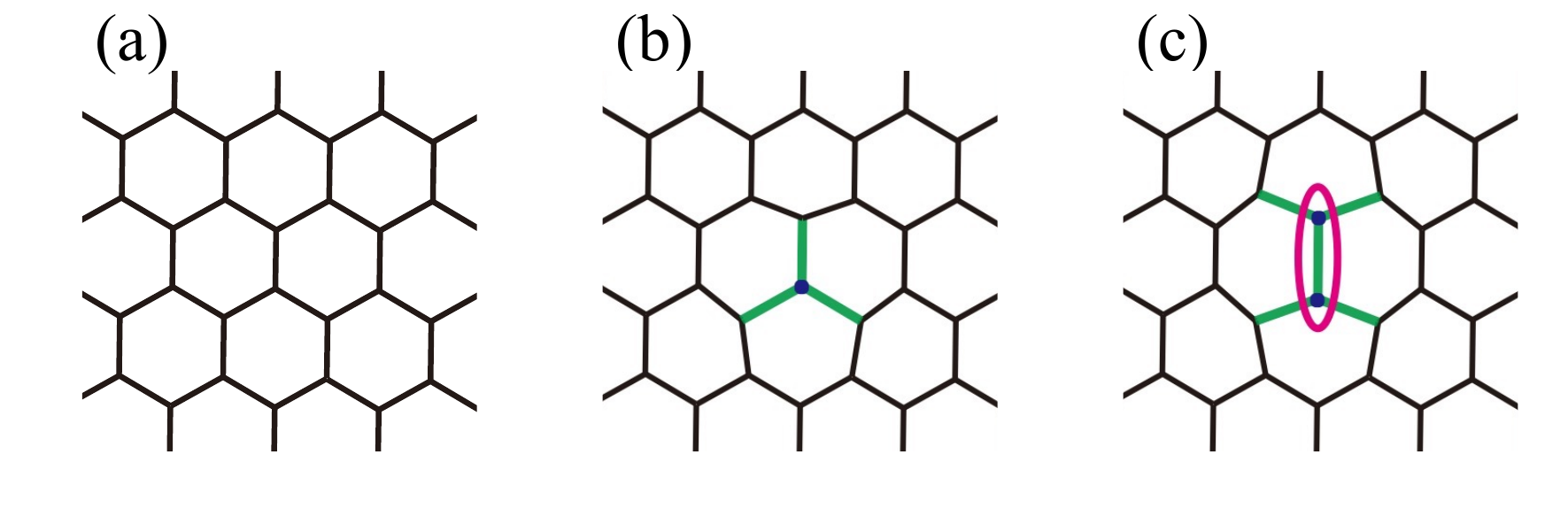}
\caption{Illustration of bipolaron formation in a honeycomb lattice with strong covalent bonds (a).
    (b) Because a doped carrier weakens the strength of a covalent bond by one half, in strongly covalent systems the carriers are therefore dressed by surrounding lattice deformation, forming so-called polarons.
    (c) These polarons have a strong tendency toward binding into bipolarons (represented by the pink ellipsoid) to reduce the number of weakened bonds, from 6 to 5.
    These bond-centered bipolarons naturally form a type of emergent Bose liquid.
}\label{Sfig4}
\end{figure}

As outlined in the main text, the condition for the EBL to form is quite general.
It assumes an emergent binding between itinerant carriers at the energy scale higher than the temperature and energy scale of experimental observations.
Here, as an elucidating example, we present one such binding mechanism through the formation of “bipolaron''~\cite{Alexandrov,Jun_1992} in strong covalent bond materials like graphene.
In systems with robust covalent bonds, the emergence of bipolarons, a distinctive variant of emergent Bose liquid, should not come as an unexpected occurrence at a considerably high energy scale.
Given that the energy reduction of a covalent bond in these systems is mostly through doubly occupying the bonding orbital, it is easy to see that a doped carrier would weaken the covalent bond by half of its strength. For extremely strong covalent systems, this weakening would be of rather large energy as well, causing a significant bond stretching and associated with it strong deviation of atomic position.
As shown in Fig.~\ref{Sfig4}(b), the carriers are therefore heavily “dressed'' by such lattice distortion and become the so-called “polaron''~\cite{emin_2012}.
Furthermore, as shown in Fig.~\ref{Sfig4}(c), such polarons can often lower the system energy if two of them bind together and form a bipolaron, since in this case, they disrupt one less (5 instead of 6) bonds than if they were separated.

Again, for the main conclusions of our manuscript, the specific details of the binding mechanism are actually not very important, since lower-energy physics are unable to overcome the binding scale, and are therefore \textit{completely insensitive} to it.
This is analogous to the well-known example of an ideal gas consisting of H$_2$ molecules, whose properties at room temperature and ambient pressure are completely insensitive to the higher-energy microscopic mechanism that binds the two H atoms into a molecule.

\end{appendices}

\bibliography{sn-bibliography}% 
\end{document}